# Visualization of Oxygen Vacancies and Self-doped Ligand Holes in La$_3$Ni$_2$O$_{7-\delta}$

Zehao Dong[1,†], Mengwu Huo[2,3,†], Jie Li[4,†], Jingyuan Li[2,3], Pengcheng Li[5], Hualei Sun[3,6], Yi Lu[4,7,*], Meng Wang[2,3,*], Yayu Wang[1,8,9,*], and Zhen Chen[10,11,*]

[1]State Key Laboratory of Low Dimensional Quantum Physics, Department of Physics, Tsinghua University, Beijing 100084, China
[2]Center for Neutron Science and Technology, School of Physics, Sun Yat-Sen University, Guangzhou, Guangdong 510275, China
[3]Guangdong Provincial Key Laboratory of Magnetoelectric Physics and Devices, Sun Yat-Sen University, Guangzhou, Guangdong 510275, China
[4]National Laboratory of Solid State Microstructures and Department of Physics, Nanjing University, Nanjing 210093, China
[5]School of Materials Science and Engineering, Tsinghua University, Beijing 100084, P. R. China
[6]School of Science, Sun Yat-Sen University, Shenzhen, Guangdong 518107, China
[7]Collaborative Innovation Center of Advanced Microstructures, Nanjing University, Nanjing 210093, China
[8]New Cornerstone Science Laboratory, Frontier Science Center for Quantum Information, Beijing 100084, China
[9]Hefei National Laboratory, Hefei 230088, China
[10]Beijing National Laboratory for Condensed Matter Physics, Institute of Physics, Chinese Academy of Sciences, Beijing 100190, China
[11]School of Physical Sciences, University of Chinese Academy of Sciences, Beijing 100049, China

*Corresponding author. Email: yilu@nju.edu.cn; wangmeng5@mail.sysu.edu.cn; yayuwang@tsinghua.edu.cn; zhen.chen@iphy.ac.cn
†These authors contributed equally.

**Abstract**

The recent discovery of superconductivity in La$_3$Ni$_2$O$_{7-\delta}$ under high pressure with a transition temperature around 80 K has sparked extensive experimental and theoretical efforts. Several key questions regarding the pairing mechanism remain to be answered, such as the most relevant atomic orbitals and the role of atomic deficiencies. Here, we develop a new energy-filtered multislice electron ptychography technique, assisted with electron energy loss spectroscopy, to address these critical issues. Oxygen vacancies are directly visualized and are found to primarily occupy the inner apical sites, which have been proposed to be crucial to superconductivity. We precisely determine the nanoscale stoichiometry and its correlation to the oxygen $K$ edge spectra, which reveals a significant inhomogeneity in the oxygen content and electronic structure within the sample. The spectroscopic results also unveil that stoichiometric La$_3$Ni$_2$O$_7$ is in the charge-transfer regime, with holes that are self-doped from Ni sites into O sites. The outer apical oxygen is found to be less relevant to the low-energy physics and can be safely disregarded in theoretical models. These observations will assist in further development and understanding of superconducting nickelate materials. Our imaging technique for quantifying atomic deficiencies can also be widely applied in materials science and condensed matter physics.

**Main text**

As an analog to cuprates, nickelates have been candidates to host high-temperature superconductivity for ages[1,2]. The first nickelate superconductor, hole-doped infinite-layer RNiO$_2$ (R represents rare-earth elements), has reached a transition temperature ($T_c$) around 20 K at ambient conditions and 31 K under pressure[3–5]. Recently, high-temperature superconductivity with onset $T_c \approx 80$ K has been discovered in the nominal Ruddlesden-Popper (RP) phase compound La$_3$Ni$_2$O$_7$ under high pressure[6–8]. In La$_3$Ni$_2$O$_7$, a straightforward electron counting reveals that the $3d$ electrons of Ni form a half-filled $3d_{z^2}$ band and a quarter-filled $3d_{x^2-y^2}$ band, and a strong correlation in $3d_{z^2}$ orbitals has been identified in experiments[9,10]. Consequently, the significance of the interlayer superexchange has been proposed theoretically, which involves the virtual hopping of the strongly correlated Ni-$3d_{z^2}$ electrons through the inner apical oxygen $2p_z$ orbitals bridging the two neighboring NiO$_2$ planes in each unit cell[11–16]. Following this picture, several studies argue that a small amount of oxygen deficiencies on this site could suppress superconductivity[17,18].

Experimentally, however, it has been challenging to characterize such light atoms like O, especially to achieve quantitative analysis for its vacancies. With scanning transmission electron microscopy (STEM), conventional method like high angle annular dark field (HAADF) imaging is unable to visualize oxygen atoms. Integrated diffrential phase contrast (iDPC)[19] and annular bright field (ABF)[5] techniques are widely adopted for oxygen imaging. However, iDPC imaging relies on the phase object approximation, which is only applicable to thin samples[20], and the response in ABF images is not linearly dependent on the number of atoms in projection[21], which hinders its application for quantifying atomic vacancies.

The recently developed phase retrieval technique, multislice electron ptychography (MEP), has been demonstrated to overcome these limitations, and even provides resolution along the depth dimension[22]. In an ideal MEP reconstruction, the phase of the complex object function is linearly proportional to the electrostatic potential from the atoms[22,23]. Therefore, MEP stands out as a tool capable of determining the nanoscale stoichiometry of crystals, but such an application has not been realized in experiments so far. The imaging condition of MEP is also compatible with the atomic-resolution spectroscopic analysis by electron energy loss spectroscopy (EELS), allowing a direct correlation of the oxygen content with the electronic structure above the Fermi energy.

Here, we use an improved approach, energy-filtered MEP, to directly visualize the distribution of oxygen vacancies in La$_3$Ni$_2$O$_{7-\delta}$, and find that they are predominantly located on the inner apical sites. The corresponding EELS results on the O-$K$ edge suggest a partially filled

O-2$p$ band with self-doped holes, which place La$_3$Ni$_2$O$_{7-\delta}$ adjacent to cuprates in the Zaanen-Sawatzky-Allen scheme[24], in stark contrast to the infinite layer nickelates[25,26]. We also find that the EELS signal is strongly modified by the concentration of oxygen vacancies, and the feature near the Fermi energy mainly stems from the inner apical and planar oxygen sites. Therefore, a minimal model for La$_3$Ni$_2$O$_7$ must include the contribution from these O-2$p$ orbitals, while the outer apical oxygen should be less relevant to the low-energy physics.

**Visualization and Statistics of the Oxygen Vacancies**

We first perform simulations to demonstrate the superiority of MEP over the other imaging techniques, namely, HAADF, iDPC, and ABF, in its capability to visualize and quantify the total number of oxygen atoms projected along a single column. The basic setup for MEP is illustrated in Fig. 1a, where a defocused electron probe is scanned along the sample, while the diffraction patterns from all scanning positions are combined to form a four-dimensional STEM (4D-STEM) dataset[22]. A typical reconstructed three-dimensional object is depicted in Fig. 1b, and the subsequent projected MEP phase image is derived by summing the phase of all slices along the depth dimension (Fig. 1d is an example). Figure 1c presents the structural model of the La$_3$Ni$_2$O$_{7-\delta}$ crystal utilized in our simulations. Specifically, the crystal is embedded with oxygen vacancies randomly distributed within the inner apical sites, one of which is shown and marked with the black dashed circle as an example. Hereafter, the images from simulations and experiments are all projected along the [110] axis of La$_3$Ni$_2$O$_{7-\delta}$, which allows for the visualization of all three inequivalent oxygen sites.

While oxygen atoms are elusive in the HAADF image (Fig. 1e), they are captured in the MEP, ABF, and iDPC images (Fig. 1d, f, g). Notably, the MEP image hosts the best resolution, which is mainly attributed to the extra information captured beyond the Abbe diffraction limit[27]. Moreover, the contrast is discernible between the oxygen atoms on the inner apical sites and the other ones, as denoted by the yellow dashed shapes in Fig. 1d, f, g. Accordingly, the content of oxygen vacancy can be evaluated. Figure 1h elucidates the correlation between the contrast of inequivalent oxygen atoms and the value of $\delta$ in our structural model using these three techniques. ABF overestimates the vacancy concentration and behaves nonlinearly, while iDPC tends to underestimate the phase contrast due to the extended tail from La atoms. Notably, the contrast in ABF and iDPC is sensitive to the imperfection of experimental conditions, such as the small mis-tilt of samples, the geometry of detectors, and the deviation from the ideal focal plane, imposing significant errors in quantification[21,28]. On the contrary, these factors can be corrected during MEP reconstructions, which is the only technique exhibiting a linear phase response and an outstanding accuracy. Therefore, MEP is confirmed to be the ideal methodology to determine the nanoscale stoichiometry of crystalline materials like La$_3$Ni$_2$O$_{7-\delta}$. This is also supported by the corresponding experimental results (Fig. S1), where the oxygen atoms are only captured with sufficient contrast in MEP results.

Having established the linear relationship between the phase value and the oxygen concentrations within an atomic column, we turn to the experimental results on imaging the oxygen vacancies in La$_3$Ni$_2$O$_{7-\delta}$. Energy-filtered 4D-STEM[29] datasets were used for MEP to minimize the inelastically scattered electrons that incoherently contribute to the diffraction patterns and degrade the quality of the reconstructed image. Figure 2a presents a sample slice taken at a depth of 6.6 nm, wherein a missing inner apical oxygen (marked with a yellow dashed circle) can be identified. This is more clearly visualized by the linecuts along the two dashed arrows (Fig. 2b), where the phase of the inner apical O is much weaker in the blue curve. The existence of oxygen vacancies is also confirmed in the third dimension, as illustrated by the depth profile along the blue dashed arrow in Fig. 2c. A phase dip at the inner apical site, centered around a depth of 6.6 nm, is consistent with local oxygen vacancies. Hence, the MEP technique indeed captures oxygen vacancies in all three dimensions (A full image stack is provided in Fig. S2).

Next, we move on to estimate the nanoscale concentration of the oxygen vacancies, which turns out to host significant inhomogeneity within the same sample, exemplified by the three representative regions showcased in Fig. 2d-f (see Fig. S3 to Fig. S5 for complete datasets). The three presented images are projected and summed along the depth dimension. The inner apical O sites in Fig. 2d are nearly identical in phase compared to the other O sites. Notably, however, the oxygen atoms marked within the yellow dashed rectangles in Fig. 2e exhibit a weaker phase than the other ones. In Fig. 2f, a weaker phase can be observed for nearly all inner apical oxygens. Therefore, the oxygen vacancies predominantly occupy the inner apical site, which is consistent with earlier model refinements using neutron diffraction[30]. These results also unveil a variation in oxygen concentration among these regions, which are quantitatively evaluated and depicted in Fig. 2g-i. For each region, we count the phase of three distinct O sites: the outer apical ones, the inner apical ones, and the planar ones. From Fig. 2g to Fig. 2i, the average phase of the inner apical O shifts to lower values, while the phase of the other O sites remains nearly constant. Correspondingly, the estimated values of $\delta$ amount to $0.04 \pm 0.11$, $0.17 \pm 0.12$, and $0.34 \pm 0.22$, respectively. This signifies substantial inhomogeneity in oxygen vacancy distribution within this material. Notably, for the region with more oxygen vacancies, the variation in oxygen content is also larger, as indicated by the differences in the value of standard deviation among these three regions. With the increase in oxygen vacancies, we also observe an extension in bond length between Ni and the outer apical O (Fig. S6).

**O $K$-edge EELS and the Self-doped Ligand Holes of $La_3Ni_2O_{7-\delta}$**

Having determined the nanoscale stoichiometry, we shift to the oxygen $K$-edge EELS on these regions, where an O-$1s$ core electron (with binding energy around 530 eV) is excited to above the Fermi energy. Similar to X-ray absorption spectroscopy (XAS), dipole selection rules restrict the angular momentum transfer $\Delta l = \pm 1$, thus the O-$K$ near edge structure correlates with the unoccupied density of states with O-$2p$ characters, i.e. the ligand hole states[31]. As shown in Fig. 3a, for the region with negligible oxygen vacancies, the EELS exhibits a strong pre-edge peak at ~528 eV and a broad shoulder around ~533 eV, before reaching the maximum at ~535 eV. The dominant peak at 535 eV and the 533-eV shoulder are regarded as the result of the transition into unoccupied O $2p$ states hybridized with empty La $5d$ states[32–34], which hardly contribute to the low-energy physics. Intriguingly, as the concentration of oxygen vacancy rises, the pre-edge peak at ~528 eV becomes weaker, and ultimately disappears when $\delta$ reaches 0.34, while the other features remain almost the same. Hence, the 528-eV peak is directly correlated to the concentration of inner apical oxygen vacancies. Accordingly, the spatial distribution of the oxygen vacancies can be visualized through a large-area O-$K$ edge mapping (Fig. S7), where the pre-edge feature is found to be inhomogeneous within a length scale of ~40 nm.

Our EELS results are in stark contrast to those from the infinite-layer nickelate parent compounds $NdNiO_2$ and $LaNiO_2$, where both EELS and XAS show hardly any pre-edge features[25,26]. This is expected, since the formal valence of Ni at 2.5+ in $La_3Ni_2O_7$ is much higher than the Ni valence at 1+ in $RNiO_2$ compounds, indicating a substantial reduction in the charge transfer energy. Moreover, the presence of a strong pre-edge peak at 528 eV is an indication of a partially filled O-$2p$ band[35,36], as illustrated in Fig. 3b, where holes are self-doped from Ni sites into the O sites in the ground state. Therefore, $La_3Ni_2O_7$ is in the charge-transfer regime rather than the Mott-Hubbard regime, in agreement with several theoretical studies[37–39]. Our DFT calculation of partial density of states (PDOS) for different atoms also suggests a similar arrangement of energy bands (Fig. 3c), where a finite oxygen PDOS exists near the Fermi energy due to its hybridization with Ni-$3d$ orbitals. In this picture, the disappearance of the 528-eV pre-edge peak could be explained by the filling up of the O-$2p$ band, since each oxygen vacancy donates two electrons into the system. Notably, the upper Hubbard band (UHB) of Ni-$3d$ electrons is not resolved in our experiments, which is likely concealed within the La $5d$ states.

Making use of the high spatial resolution of STEM, we extend the EELS study to the atomic scale, from which the pre-edge features can be well separated for the inequivalent oxygen sites (Fig. 4a). As denoted by the blue and green arrows, the peak at 528 eV originates predominantly from the inner apical and planar sites, which is also captured by DFT calculations shown in Fig. 4b. This feature is suppressed on the outer apical oxygen sites due to a larger Ni-O distance and a weaker hopping integral, which could significantly reduce the $p$-$d$ hybridization. On the other hand, the 533-eV shoulder is mainly contributed by the outer apical oxygen, as marked by the orange arrow in Fig. 4a. Compared to the DFT calculation results in Fig. 4b, the broad shoulder is likely a result of the outer apical $p_{x,y}$ orbitals hybridized with empty La-$5d$ orbitals. These observations suggest that the outer apical oxygen is less relevant to the low-energy physics of $La_3Ni_2O_7$ and could be safely disregarded in a minimal model for the superconductivity.

Consequently, $La_3Ni_2O_7$ can be simplified into the minimal model illustrated in Fig. 4c, where its transport properties originate from the two $NiO_2$ planes and the bridging inner apical oxygens. Specifically, the oxygen PDOS near the Fermi energy primarily comes from the $p_z$ orbital of inner apical O and the $p_{x,y}$ orbitals of planar O (Fig. S8). Together with the $d_{x2-y2}$ and $d_{z2}$ orbitals of Ni, the main physical properties of $La_3Ni_2O_7$ should lie within the bilayer model schematically shown in Fig. 4c.

**Discussion**

Our observation of ligand holes resembles the case of overdoped cuprate superconductors, wherein a similar pre-edge peak is found at a similar energy[40,41], whose spectral weight is transferred from the UHB. In the case of $La_3Ni_2O_7$, however, holes are self-doped into the ligand sites due to the intrinsic arrangement of energy levels, without externally introduced dopants. As a result, the valence of Ni should be lower than 2.5+ and closer to the stable 2+ valence state. We can also provide some insights into the superconducting stoichiometric phase from the electronic structure derived from our combined MEP and EELS results. The candidate could hardly be a phase with $\delta > 0.34$, in which case the pre-edge peak in EELS has already disappeared, and the carrier density should be significantly reduced. Additionally, the presence and uneven distribution of O vacancies might be a critical factor for the low superconducting volume fraction in susceptibility measurements under high pressure[42] and the inhomogeneity in transport measurements[7].

In summary, we present a combined study of energy-filtered MEP and EELS to unveil the role of ligand oxygen in $La_3Ni_2O_{7-\delta}$. The off-stoichiometry oxygen vacancies mainly reside on the inner apical sites, and exhibit a considerable spatial fluctuation which could be responsible for the inhomogeneity found in the previous transport and magnetic measurements. The electronic structure, derived from EELS, reveals that $La_3Ni_2O_7$ is in the charge-transfer regime with self-doped ligand holes. A strong *p-d* hybridization is observed in the 2*p* orbitals of inner apical and planar oxygen atoms. The outer apical oxygen, however, is less relevant to the low-energy physics and could be neglected in the minimal theoretical models. Further theoretical and experimental studies are required to determine the possible pairing mechanism in its superconducting phase under pressure. Moreover, the capability of MEP to determine the content of atomic deficiencies paves way to a wide range of applications in materials science.

## Acknowledgements

We thank Guang-Ming Zhang for helpful discussions and Tieqiao Zhang for assistance during the experiments. This work was supported by the National Key Research and Development Program of China (MOST) (Grant No. 2023YFA1406400, No. 2023YFA1406500, No. 2022YFA1403000, and No. 2023YFA1406002), the National Natural Science Foundation of China (Grant No. U22A6005, No. 52273227, No. 12174454, and No. 12274207), the Basic Science Center Project of NSFC (No. 52388201), the Innovation Program for Quantum Science and Technology (No. 2021ZD0302502), the Basic and Applied Basic Research Major Programme of Guangdong Province, China (Grant No. 2021B0301030003 and No. 2021B1515120015), the Guangzhou Basic and Applied Basic Research Funds (Grant No. 202201011123), and the Guangdong Provincial Key Laboratory of Magnetoelectric Physics and Devices (Grant No. 2022B1212010008). Y.W. is partially supported by the New Cornerstone Science Foundation through the New Cornerstone Investigator Program and the XPLORER PRIZE. This work used the facilities of the National Center for Electron Microscopy in Beijing at Tsinghua University.

## Author contributions

Z.C., Y.W., Y.L. and M.W. supervised the research. Z.D. performed the simulations and the STEM experiments. P.L. contributed to the setup of energy-filtered 4D-STEM from K3 camera. M.H., J.L., H.S. and M.W. performed the single crystal growth and transport measurements. J.L. and Y.L. performed the DFT calculations. Z.D., Y.W. and Z.C. wrote the paper with inputs from all authors.

## Competing interests

The authors declare no competing interests.

## Additional information

Correspondence and requests for materials should be addressed to the corresponding authors.

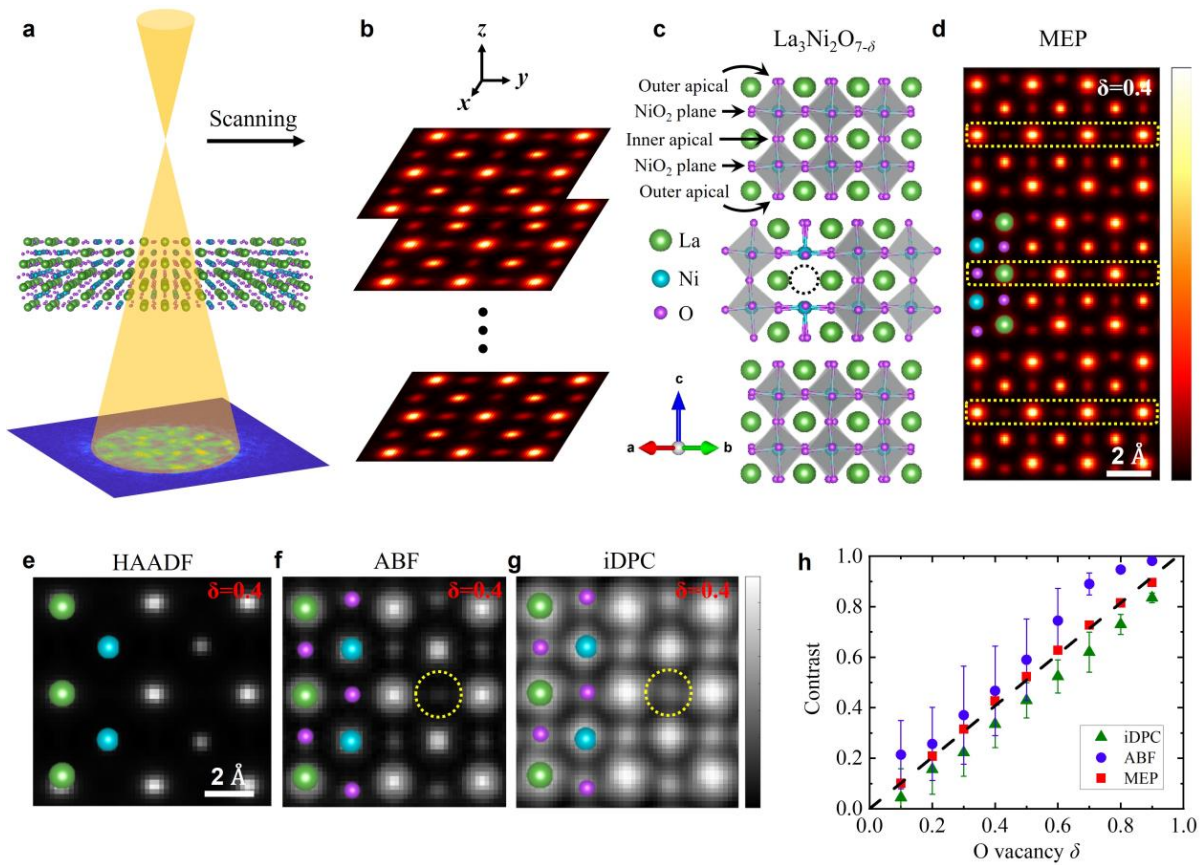

**Figure 1| The experimental setup for MEP and simulations in determining the O content. a**, Illustration of the experimental setup for MEP, where an overfocused probe is scanned along the sample, and the diffraction pattern at each position is captured. **b,** A schematic of the reconstructed three-dimensional object with MEP. **c,** A structural model for $La_3Ni_2O_{7-\delta}$ projected along the [110] axis, with one oxygen vacancy on the inner apical site as an example, which is marked with the black dashed circle. Three of the inequivalent O sites are denoted as the outer apical O, the $NiO_2$ plane (planar) O, and the inner apical O, respectively. **d,** The simulated projected MEP image for $\delta = 0.4$, where all of the atoms are visualized with high resolution. Inner apical oxygens with lower phases are marked with the yellow dashed rectangles, as an indication of vacancies. **e-f,** The simulated HAADF (**e**), ABF (**f**), and iDPC (**g**) images for $\delta = 0.4$. While the HAADF image only captures La and Ni atoms, O atoms are visualized in the ABF and iDPC images. The vacancies on the inner apical site can also be identified and marked with the dashed circles. The contrast of ABF image is reversed for clarity. **h,** The relationship between the measured phase contrast of different oxygen atoms and the vacancy concentration given in the structure model, using MEP, iDPC, and ABF, respectively. The black dashed line is a linear fit of the MEP results.

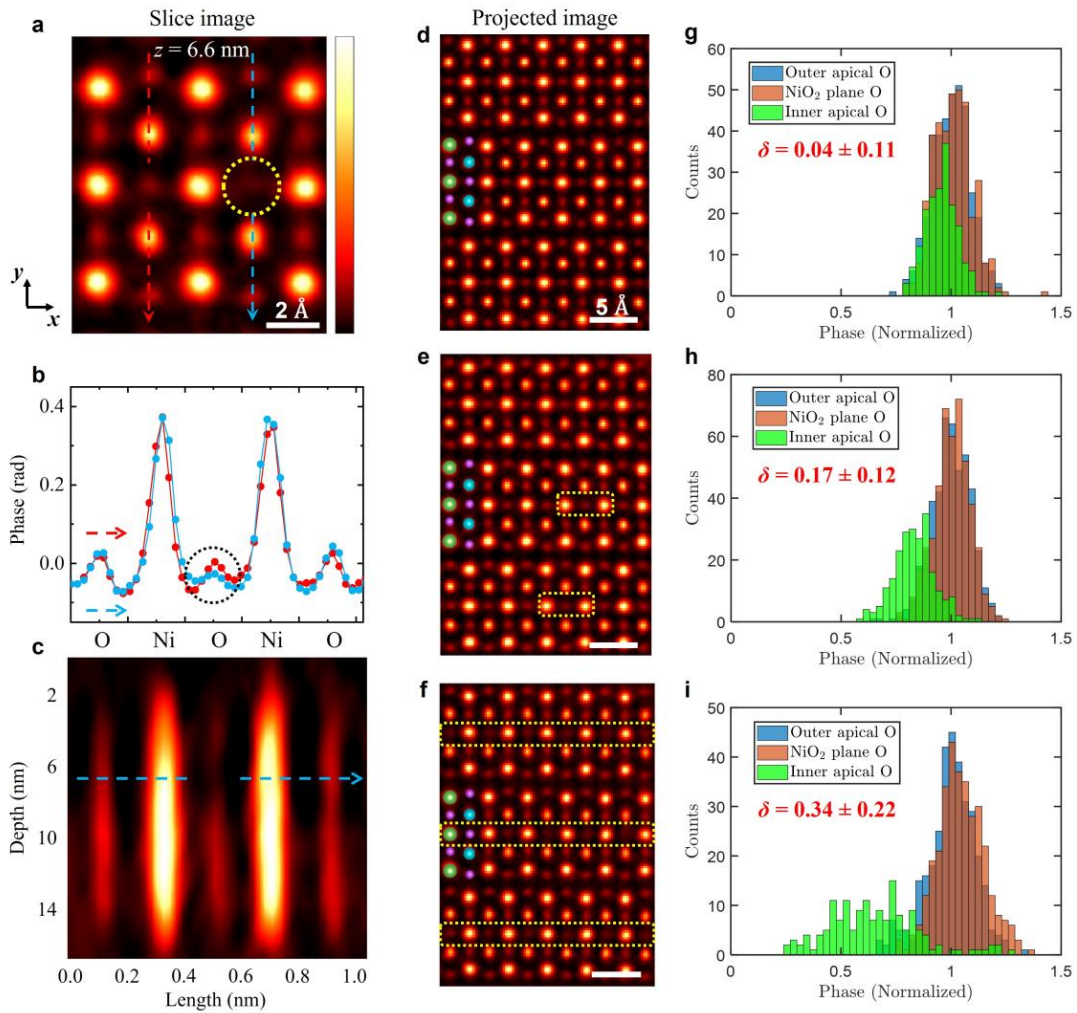

**Figure 2| Experimental visualization and statistics of the oxygen vacancies in $La_3Ni_2O_{7-\delta}$. a,** Phase image of a slice from the depth of 6.6 nm in the MEP reconstructed result, where a missing oxygen atom on the inner apical site is marked with the yellow dashed circle. **b,** Two linecuts along the dashed arrows in **a**, and the directions are illustrated as the short-dashed arrows in **b**. The arrows in **a** are broken in the middle to allow for a direct visualization of the inner apical O atoms. The black dashed circle marks the main difference from these curves, where a phase difference exists on the inner apical site. **c,** The depth profile along the blue dashed arrow in **a**. Ni and O atoms are annotated for both **b** and **c**. **d-f,** The projected MEP phase images from three regions with different oxygen vacancy concentrations, where the yellow dashed rectangles select several evident vacancies, along with the adjacent La atoms. **g-i,** the histograms of the phases from distinct oxygen sites, corresponding the regions from **d-f**, respectively. The average phase for the outer apical oxygens is normalized to one. Each histogram counts atoms within a total volume of approximately 10×10×10 $nm^3$ including about 1000 oxygen columns. The estimated $\delta$ is labeled in each figure.

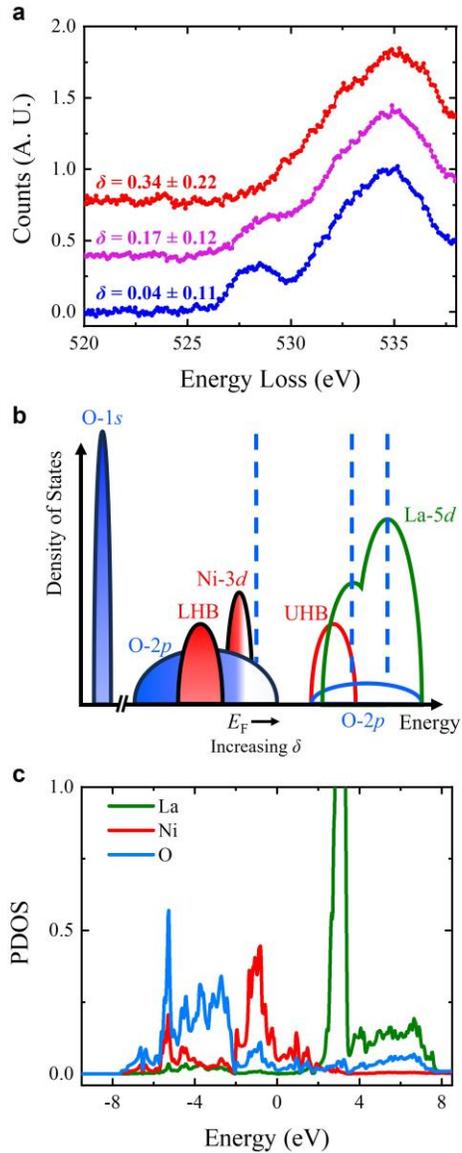

**Figure 3| O-*K* edge EELS and the electronic structure for La$_3$Ni$_2$O$_{7-\delta}$. a**, EELS taken at the sample regions with various oxygen vacancy concentrations. **b**, Illustration of the electronic structure of La$_3$Ni$_2$O$_7$, derived from the EELS results. O bands, Ni bands and La bands are marked with blue, red and green colors, respectively. Specifically, the O-2*p* orbitals may hybridize with high-energy *d* orbitals of Ni and La, and such a component is marked with the blue arc within each band. The blue dashed lines correspond to the processes that contribute to the O-*K* near edge spectra. LHB and UHB refer to the lower Hubbard band and the upper Hubbard band, respectively. **c**, DFT calculated PDOS for La, Ni, O atoms, respectively, approximately in agreement with the schematic illustration in **b**.

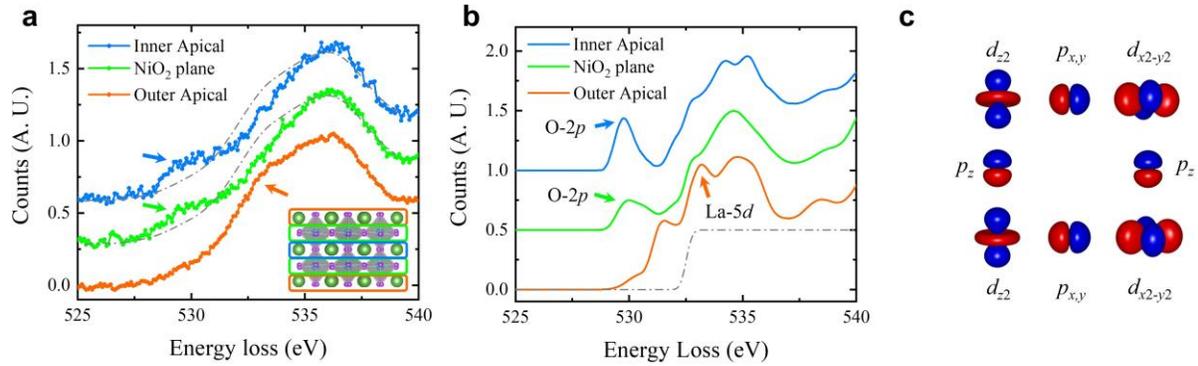

**Figure 4| The atomic-plane-resolved EELS and the relevant orbitals in $La_3Ni_2O_{7-\delta}$. a**, EELS on different oxygen atomic sites for $La_3Ni_2O_{7-\delta}$. The pre-edge features are denoted with arrows of the same color. The inset shows schematically how these spectra are obtained by summing within the selected regions. For comparison, the outer apical spectrum is copied and shifted vertically as the gray dashed lines. **b**, DFT calculated EELS spectra for the three distinct oxygen sites. In order to make a direct comparison to experiments, we add to each spectra a step function centered at 532.5 eV to simulate the absorption edge of O-*K*, which is shown as the gray dashed line. The components of the EELS features are also annotated. **c**, Schematic illustration of the inner apical O-$p_z$, planar O-$p_{x,y}$, Ni-$d_{x2-y2}$, and Ni-$d_{z2}$ orbitals that are relevant to the low-energy electronic structure in $La_3Ni_2O_{7-\delta}$, as suggested by our EELS results. The colors of the atomic orbital correspond to the signs of the wavefunctions.

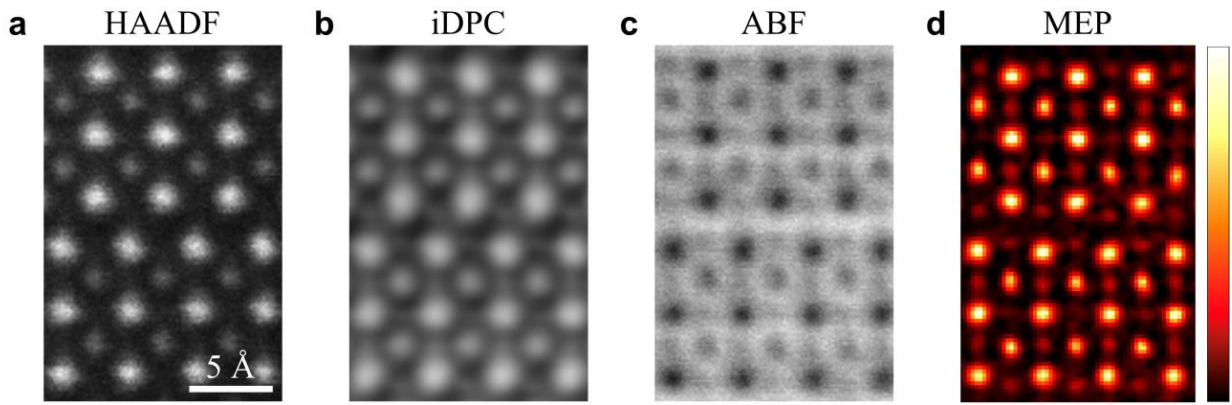

**Figure S1| Comparison of experimental results utilizing HAADF, iDPC, ABF and MEP. a-d**, Experimental imaging results on the same region from $La_3Ni_2O_{7-\delta}$, by HAADF (**a**), iDPC (**b**), ABF (**c**), and MEP (**d**), respectively. Consistent to the simulation results in the main text, the oxygen atoms are only visible in iDPC, ABF, and MEP results. However, the oxygen atoms in the iDPC and ABF images are blurry and thus difficult to quantify. Only in the MEP image are the oxygen atoms captured clearly.

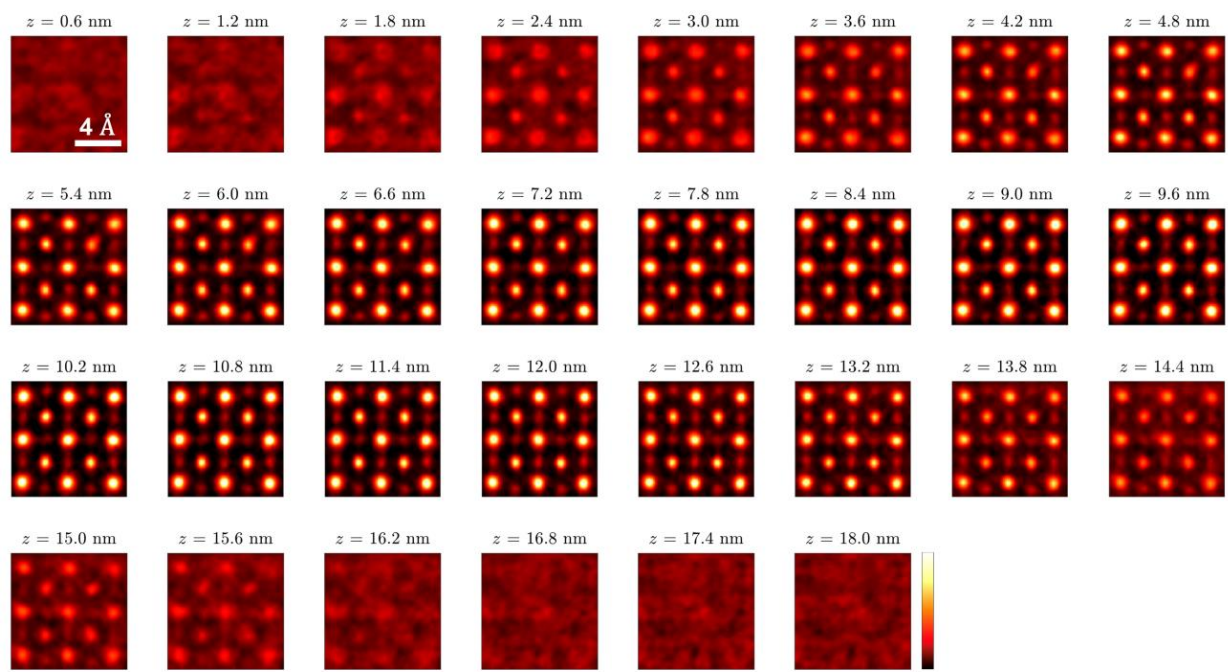

**Figure S2| A full three-dimensional reconstructed result showcasing inner apical oxygen vacancies.** The corresponding depth is annotated above each slice.

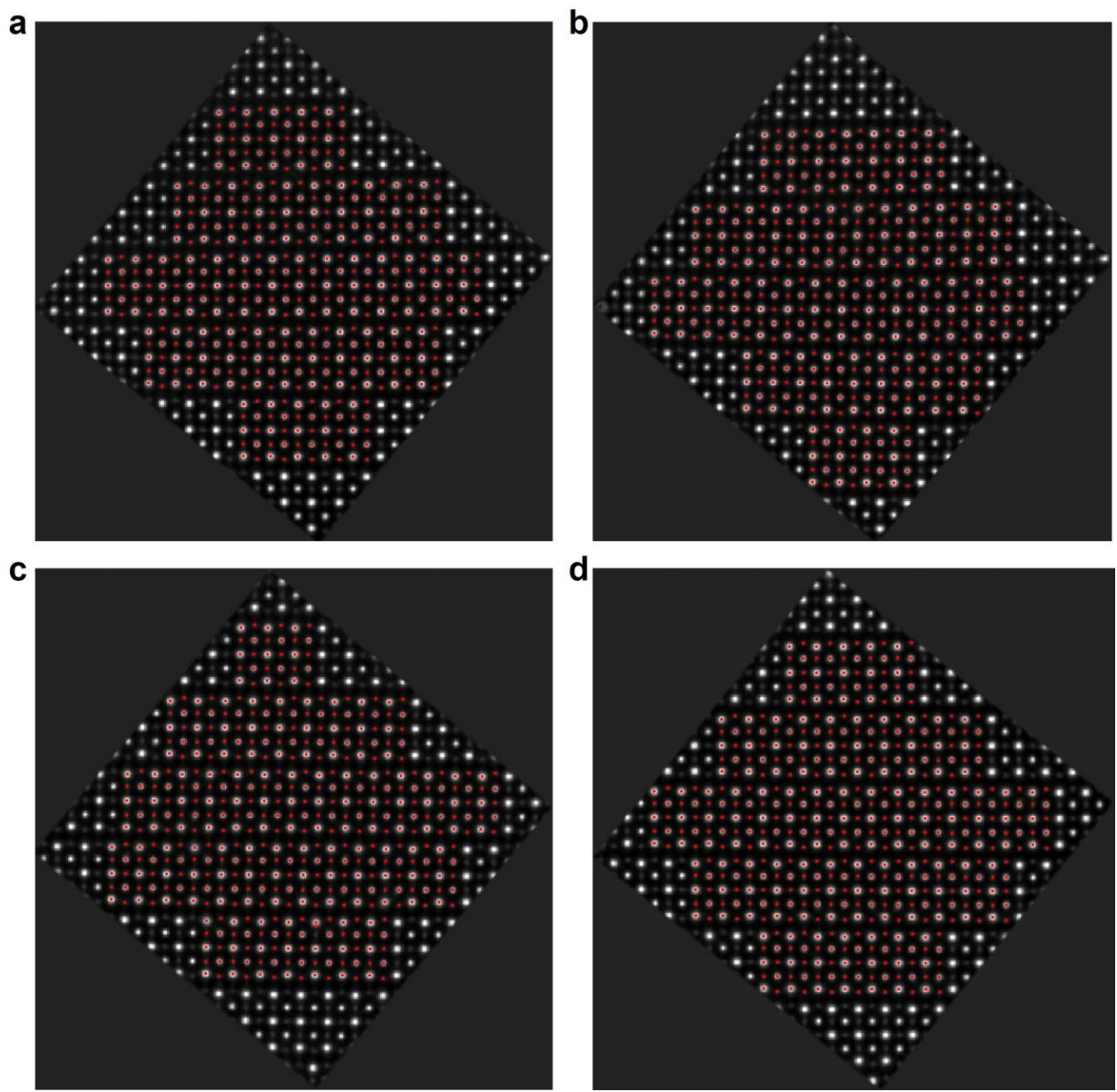

**Figure S3| The atoms for statistics in the region with $\delta = 0.04$. a-d**, Projected MEP images with atoms fitted by two-dimensional Gaussian functions, the center of each denoted by a red point. The images are presented in the gray scale.

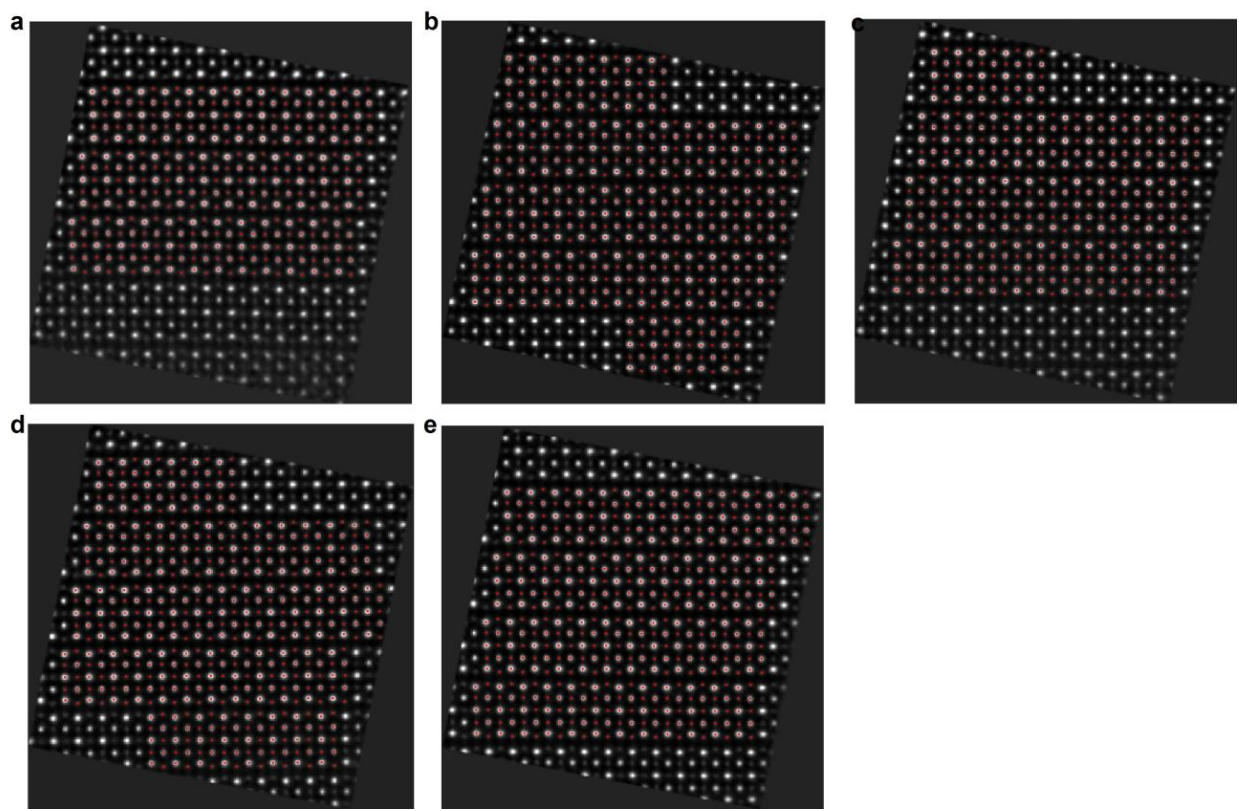

**Figure S4| The atoms for statistics in the region with δ = 0.17. a-e**, Projected MEP images with atoms fitted by two-dimensional Gaussian functions, the center of each denoted by a red point. The images are presented in the gray scale.

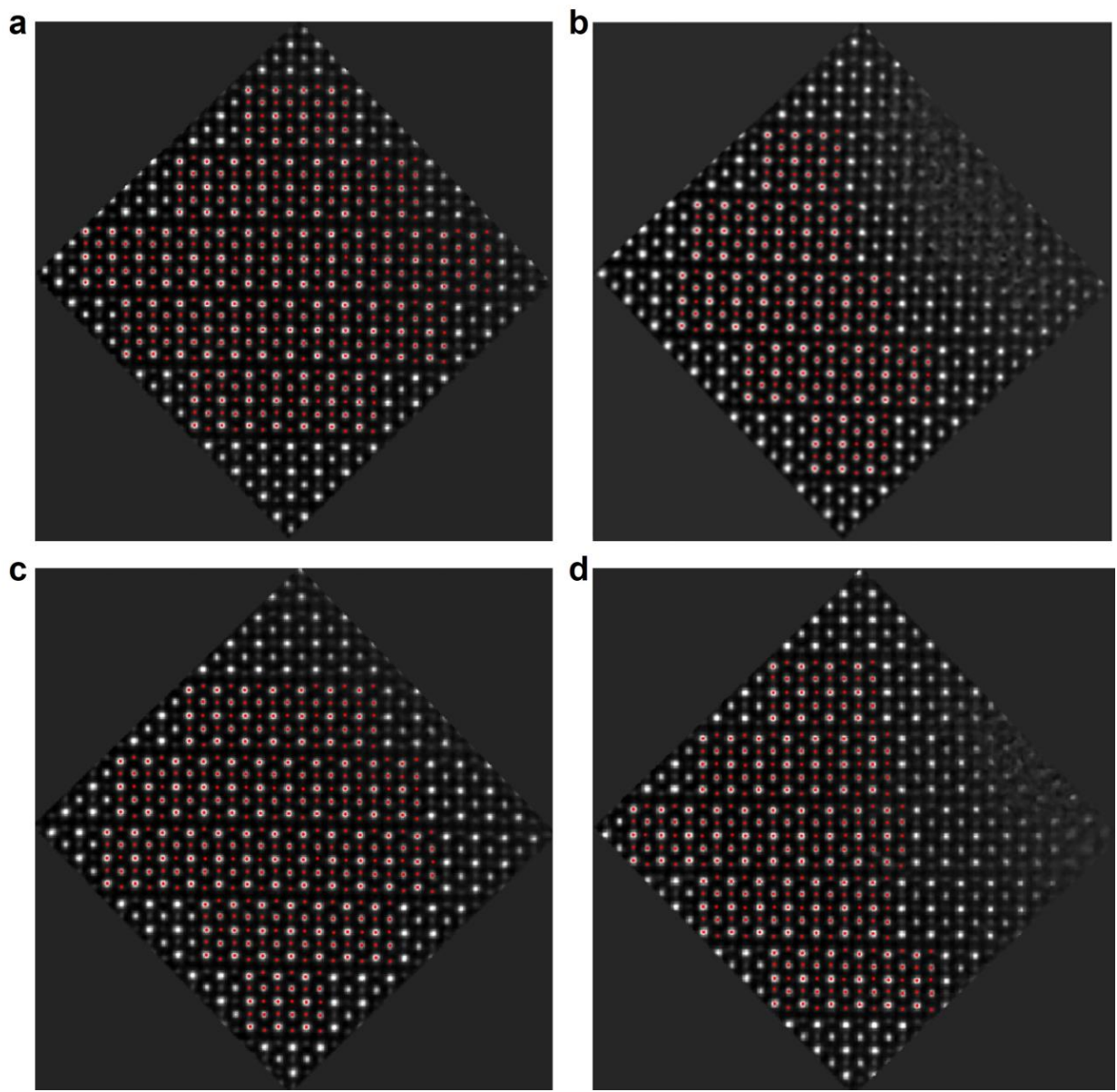

**Figure S5| The atoms for statistics in the region with $\delta$ = 0.34. a-d**, Projected MEP images with atoms fitted by two-dimensional Gaussian functions, the center of each denoted by a red point. The images are presented in the gray scale.

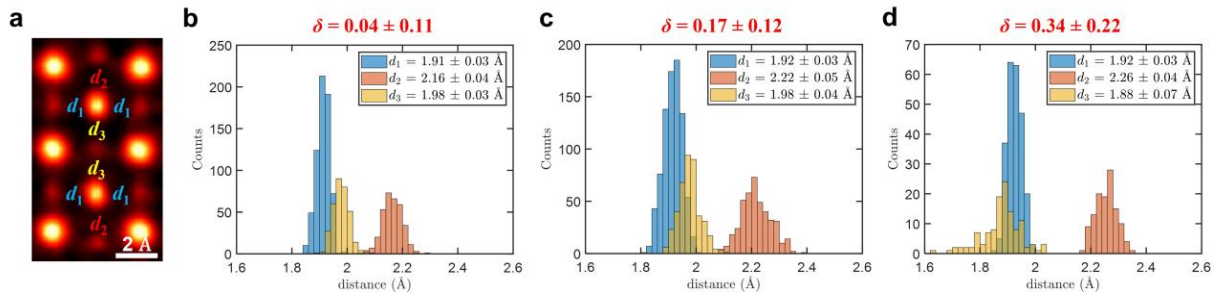

**Figure S6| Statistics for Ni-O bond lengths. a**, A schematic image illustrating the definition of three inequivalent bond lengths, $d_1$, $d_2$, and $d_3$, which corresponds to the distance between Ni and planar O, outer apical O, and inner apical O, respectively. **b-d**, The statistics of Ni-O bond lengths for the three regions in the main text, namely, regions with $\delta = 0.04$ (**b**), 0.17 (**c**) and 0.34 (**d**). The mean values and standard deviations of bond lengths are shown in the legends. The bond length $d_2$ increases as the concentration of oxygen vacancy increases, while $d_1$ remains nearly constant. The variation in $d_3$ is mainly attributed to the randomness on the inner apical sites.

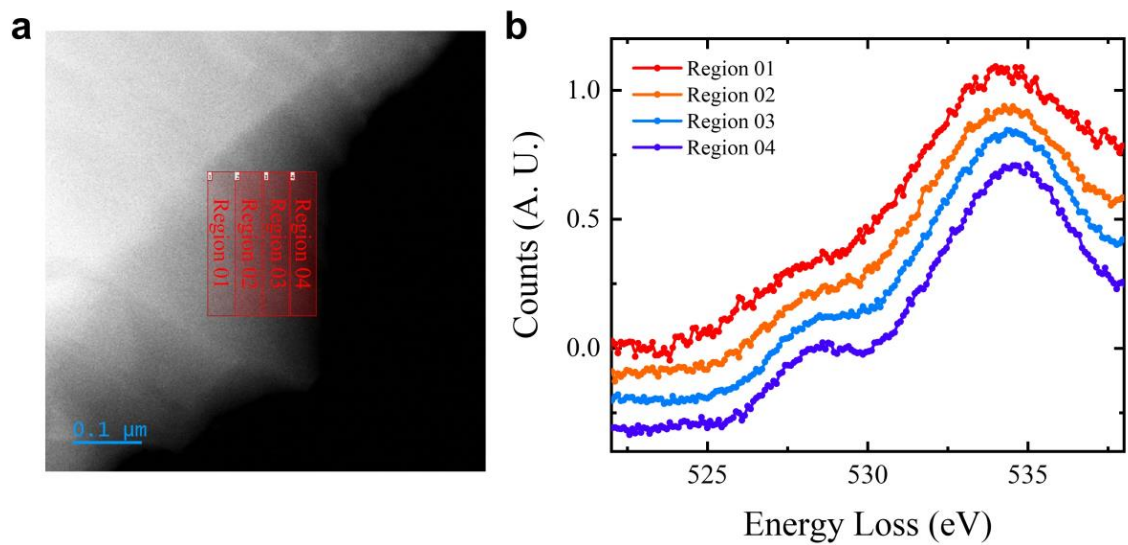

**Figure S7| The inhomogeneity of oxygen vacancies visualized by O-*K* edge EELS. a**, Large-area HAADF image of the sample. **b**, The EELS taken at the three regions selected in **a**, where the intensity of the 528-eV pre-edge peak is decreased from region 01 to region 04, each separated by a length of 40 nm. Therefore, the inhomogeneity of the oxygen vacancy is on the length scale of ~40 nm.

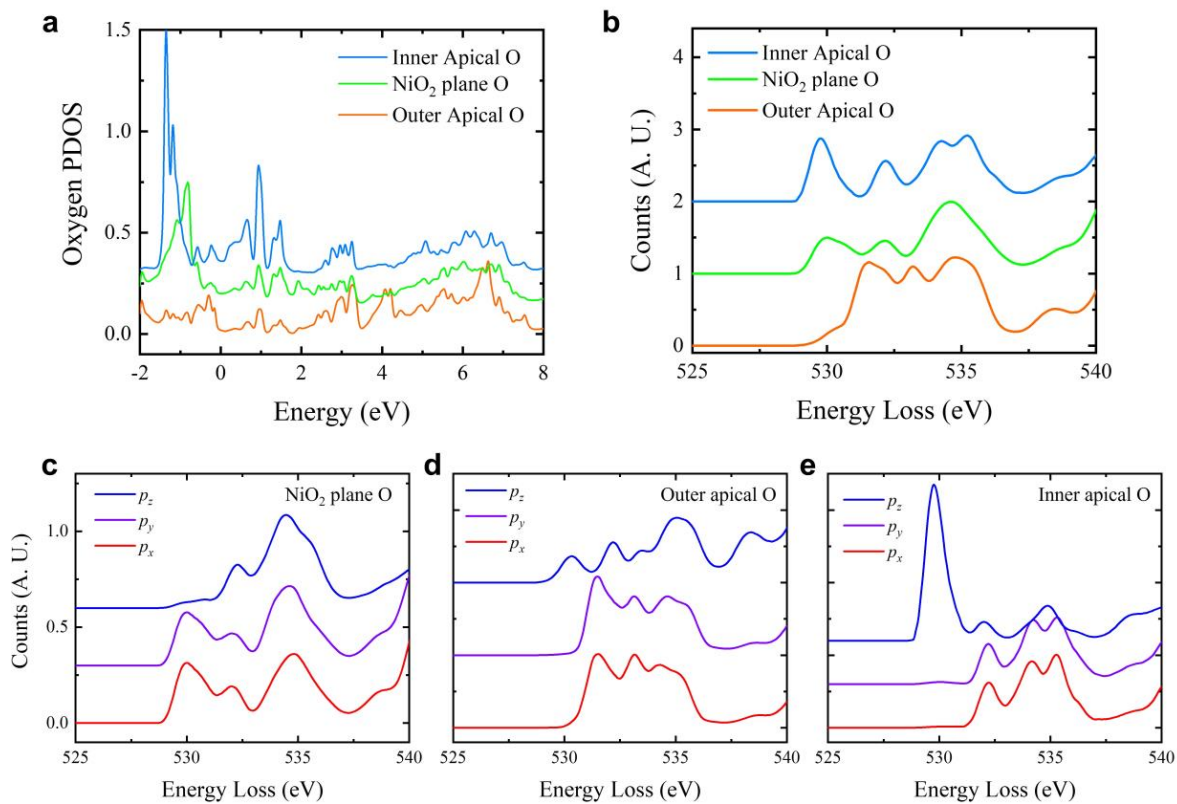

**Figure S8| DFT calculation results for La$_3$Ni$_2$O$_7$. a**, Oxygen partial density of states (PDOS) of La$_3$Ni$_2$O$_7$, which is calculated for the three inequivalent sites respectively. **b**, The corresponding EELS spectra calculated, taking into account the experimental blurring of $\Delta E$=0.6 eV. **c-e**, EELS decomposed into orbitals $p_x$, $p_y$, and $p_z$ for planar sites (**c**), outer apical sites (**d**), and inner apical sites (**e**).

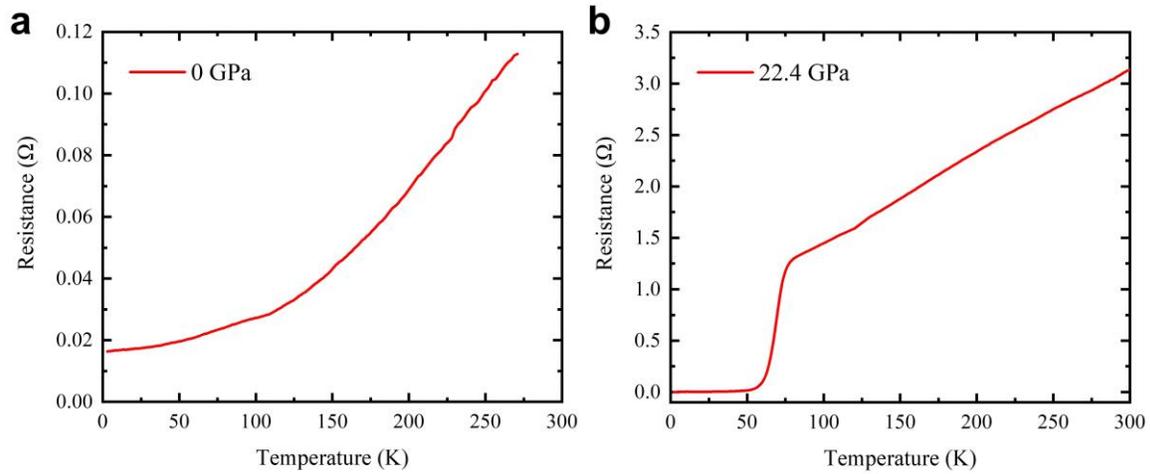

**Figure S9| Transport characterization of the single crystal used in our measurements. a**, Resistance of the $La_3Ni_2O_{7-\delta}$ single crystal versus temperature under ambient pressure. **b**, Resistance of the single crystal $La_3Ni_2O_{7-\delta}$ versus temperature under high pressure at 22.4 GPa, exhibiting a superconducting transition.

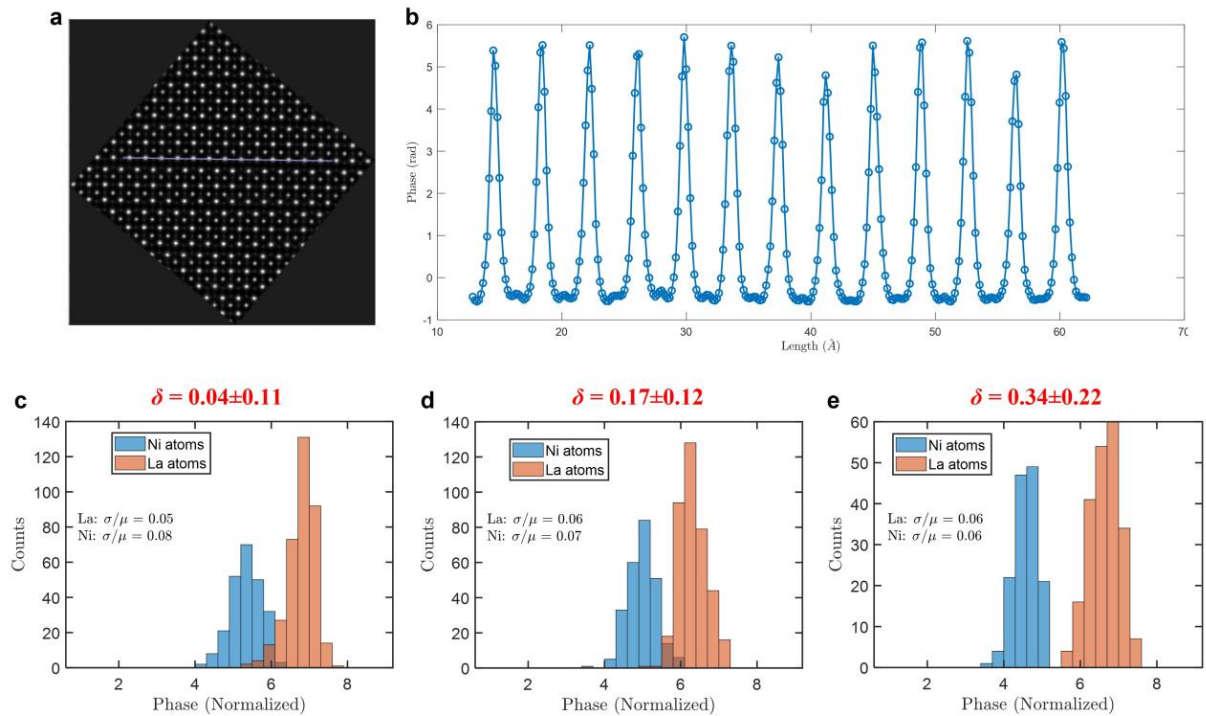

**Figure S10| Statistics for La and Ni atoms and the estimation for uncertainty. a**, A projected MEP image presented in the gray scale. **b**, The linecut along the blue curve in (**a**). The peaks correspond to La atoms, which is assumed to be free from defects. The spatial fluctuation along the line represents the uncertainty in phase. **c-e**, The histogram for the phases of La and Ni atoms from the region with $\delta = 0.04 \pm 0.11$ (**c**), $\delta = 0.17 \pm 0.12$ (**d**) and $\delta = 0.34 \pm 0.22$ (**e**), respectively. The phases for La and Ni are normalized to the outer apical oxygens, as described in Methods. As an evaluation of the precision in phase, the ratio of the standard deviation ($\sigma$) and the mean value ($\mu$) for each atom is annotated in the histogram as well. The estimated uncertainty in phase is around 6%.